\documentclass{JHEP3} 
\usepackage{amsmath}

\title{$k$-string tensions in the 4-d SU($N$)-inspired dual Abelian--Higgs-type theory}

\author{Dmitri Antonov
        \thanks{Permanent address: 
        ITEP, B. Cheremushkinskaya 25, RU-117 218 Moscow, Russia.}\\ 
        Institute of Physics, Humboldt University of Berlin\\
        Newtonstr. 15, 12489 Berlin, Germany \\       
        E-mail: \email{antonov@physik.hu-berlin.de}}

\author{Luigi Del Debbio \\
        CERN, Department of Physics, TH Division \\
        CH-1211 Geneva 23, Switzerland \\
        E-mail: \email{luigi.del.debbio@cern.ch}}

\author{Dietmar Ebert \\
        Institute of Physics, Humboldt University of Berlin\\
        Newtonstr. 15, 12489 Berlin, Germany \\       
        E-mail: \email{debert@physik.hu-berlin.de}}

\abstract{ The $k$-string tensions are explored in the 4-d
[U(1)]${}^{N-1}$-invariant dual Abelian--Higgs-type theory. In the
London limit of this theory, the Casimir scaling is found in the
approximation when small-sized closed dual strings are
disregarded. When these strings are treated in the dilute-plasma
approximation, explicit corrections to the Casimir scaling are
found. The leading correction due to the deviation from the London
limit is also derived. Its $N$-ality dependence turns out to be the
same as that of the first non-trivial correction produced by closed
strings. It also turns out that this $N$-ality dependence coincides
with that of the leading correction to the $k$-string tension, which
emerges by way of the non-diluteness of the monopole plasma in the 3-d
SU($N$) Georgi--Glashow model. Finally, we prove that, in the latter
model, Casimir scaling holds even at monopole densities close to the
mean one, provided the string world sheet is flat.}


\preprint{HU-EP-04/59 \\
CERN-PH-TH/2004-164} 

\begin{document}

\section{Introduction}

The spectrum of $k$-string tensions has been extensively explored in
recent years, both on the lattice~\cite{1}--\cite{7} and in the
continuum limit~\cite{8}--\cite{jh}.  A $k$-string is a string joining
$k$ quarks with $k$ antiquarks. Alternatively, it can be defined as a
string between sources that carry charge $k$ ($N$-ality) with respect
to the center group, $Z_N$, of the original SU($N$) group. The
knowledge of the tensions of $k$-strings, $\sigma_k$, is likely to
shed some light on the dynamics of confinement by providing some new
``phenomenological'' input that models of confinement have to satisfy,
see e.g.~\cite{DelDebbio:2002yp,Greensite:2002yn}.

Some properties of the $k$-string spectrum follow directly from first
principles. The ratio of $\sigma_k$ to the tension of the fundamental
string, $\sigma_1$, in an SU($N$) pure gauge theory is invariant under
the interchange of quarks and antiquarks, $k\leftrightarrow
(N-k)$. Moreover, in the large-$N$ limit, the interactions between
fundamental strings inside the composite $k$-string are suppressed, it
is therefore expected that in this limit their spectrum fulfils the
condition $\frac{\sigma_k}{\sigma_1} \stackrel{{k-{\rm
fixed}}\atop{N\to \infty}}{\longrightarrow} k$. Finally, the
corrections to this large-$N$ limit are expected to appear in powers
of $1/N^2$~\cite{12}.

Analytical calculations in supersymmetric
theories~\cite{Douglas:1995nw,11}, or in some approximations to
Yang--Mills theories based on the string/gauge
duality~\cite{Hanany:1997hr,Herzog:2001fq}, yield the so-called sine
scaling for the $k$-string spectrum:
$\frac{\sigma_k}{\sigma_1}=\frac{\sin\left(k
\pi/N\right)}{\sin\left(\pi/N\right)}$. In two dimensions, one can
show that the string tension ratio obeys an exact Casimir scaling:
$\frac{\sigma_k}{\sigma_1} = \frac{k(N-k)}{N-1}$. The same result is
obtained for the four-dimensional pure gauge theory, when the
computations are based either on dimensional
reduction~\cite{Ambjorn:1984dp} or on the low orders of some
perturbative expansion; see e.g. Ref.~\cite{6} for the $k$-string
spectrum in the strong coupling Hamiltonian formulation. Casimir
scaling also appears in some models of the QCD vacuum, such as the
stochastic vacuum model of QCD~\cite{8,9} and the [U(1)]${}^{N-1}$
gauge-invariant Abelian-projected theory~\cite{10}. In 3-d,
Casimir scaling with certain corrections has been
found in Ref.~\cite{jh}.

It is important to stress that no analytical computation of the
$k$-string spectrum has been done so far that could be directly
applied to the case of pure gauge theory at weak coupling; neither the
sine nor the Casimir scaling can be considered as {\it exact}\ results
in non-supersymmetric Yang--Mills theories. While both formulae yield
the correct limit as $N\to\infty$ at fixed $k$; at subleading order,
the sine scaling shows the expected $1/N^2$ corrections, whereas the
Casimir scaling leads to corrections in powers of $1/N$, thus
contradicting the results of Ref.~\cite{12}. Lattice results show that
both formulae give an approximate but satisfactory description of the
numerical data. Different lattice simulations display some discrepancy
in the spectrum, which are likely to arise from systematic errors in
the computation of the string tension. At the current level of
accuracy, lattice data do not allow a clear-cut distinction between
the two behaviors; of course there is no theoretical reason to expect
any of the two to be exact, and it would be more interesting if
lattice results could become accurate enough for one to be able to
pinpoint the parametric behavior of the subleading corrections to the
large-$N$ limit.

This paper parallels in its spirit Ref.~\cite{jh}, since it also deals
with a confining Abelian-type theory, although a 4-d one. More
specifically, we explore in this work $k$-string tensions in the 4-d
[U(1)]${}^{N-1}$ gauge-invariant Abelian-projected theory, which is
formulated in terms of dual magnetic Abelian gauge fields, neglecting
the off-diagonal degrees of freedom.  Unlike Ref.~\cite{10}, where
this was done in the Bogomol'nyi limit and on the basis of the
analysis of the classical string solutions, here we will consider the
model in the London limit (and in its vicinity), which corresponds to
an extreme type-II dual superconductor; the $k$-string tensions will
be derived from the string representation of the partition
function. Such a representation means a reformulation of the theory in
terms of the path integral over closed dual strings, which are always
present in the theory and interact with the external $k$-string. As a
brief historical remark, let us mention that, for the usual (dual)
Abelian Higgs model, the reformulation of the partition function in
terms of closed strings has been performed by many authors (e.g. in
Ref.~\cite{sr} in 4-d and in Ref.~\cite{sr0} in 3-d), in particular
with the applications to the stochastic vacuum model~\cite{SR, epj};
the Jacobian of the transformation from field to string variables has
been discussed in Ref.~\cite{sr1}; the SU($N$) generalization has been
explored, in particular for studies of the $\theta$-term~\cite{plbN}
[see Ref.~\cite{theta2} for investigations of the $\theta$-term in the
SU(2)- and SU(3)-inspired cases] and for the purposes of further
applications to the stochastic vacuum model at $N=3$~\cite{su3a,
moresu3} [see also~\cite{su3b} for related studies at $N=3$]; the
corrections emerging in the vicinity of the London limit have also
been explored~\cite{sr3}.

 The important fact is that closed strings are short-lived (virtual)
objects~\cite{22}, whose typical sizes are much smaller than both
distances between them and size of the $k$-string
world sheet. Therefore, since world-sheet tensors (also called
vorticity tensor currents) of closed strings enter the final action in
the linear combinations with the world-sheet tensor of the $k$-string,
in the leading semi-classical approximation the interaction of closed
strings with the $k$-string can be merely disregarded. This is
precisely the approximation in which the SU(2)- and SU(3)-inspired
models have been considered in Refs.~\cite{SR, epj} and~\cite{su3a},
respectively. This approximation can be further improved by treating
closed strings in the dilute-plasma approximation~\cite{moresu3}. In
this paper, specifically in Section~5, we will also account for
effects produced by the dilute plasma of closed dual strings.

The sketch of the paper is as follows. In the next section, the model
under study will be described.  In Section~3, $\sigma_k$ will be
derived in the London limit. In Section~4, we will explore corrections
to $\sigma_k$, which emerge in the vicinity of the London limit, i.e.
when Higgs bosons are not infinitely heavy. In Section~5, we will
return to the London limit and consider another type of corrections,
namely those produced by closed dual strings.  In Section~6, the main
results of the paper will be discussed once again.  In Appendix~A,
some estimates related to the main part of the text will be performed.
Finally, Appendix~B is devoted to the 3-d SU($N$) Georgi-Glashow
model; it is complimentary to Ref.~\cite{jh}, where Casimir scaling in
this model has been proved for an arbitrarily shaped surface, but at
monopole densities much lower than the mean one. Here we prove that
Casimir scaling holds already at the mean density, provided the
surface is flat.

\section{The model}

The model we are going to deal with is the generalization of the
SU(3)-inspired dual Abelian--Higgs-type theory~\cite{maedan} to the
case of arbitrary $N$~\cite{10, plbN}.  The monopole condensation is
modelled in it by the assumption that monopoles form condensates of
the dual Higgs fields.  This model can naturally be called an
effective [U(1)]${}^{N-1}$ gauge-invariant Abelian-projected
theory. Its Euclidean partition function reads:
$$
{\cal Z}_k=\int\left(\prod\limits_{i}^{} \left|\Phi_i\right| {\cal D}\left|\Phi_i\right|
{\cal D}\theta_i\right) {\cal D}{\bf B}_\mu
\delta\left(\sum\limits_{i}^{}
\theta_i\right)\exp\Biggl\{-\int d^4x\Biggl[\frac14\left({\bf F}_{\mu\nu}+{\bf F}_{\mu\nu}^k\right)^2+$$

\begin{equation}
\label{et6}
+\sum\limits_{i}^{}\left[\left|\left(\partial_\mu-
ig_m{\bf q}_i{\bf B}_\mu\right)\Phi_i\right|^2+
\lambda\left(|\Phi_i|^2-\eta^2\right)^2\right]
\Biggr]\Biggr\}.
\end{equation}
Here, the index $i$ runs from 1 to the number of positive roots ${\bf q}_i$ of the SU($N$)-group, that is $N(N-1)/2$.
Next, $g_m$ is the magnetic
coupling constant related to the
electric one, $g$, by means of the Dirac
quantization condition $g_mg=4\pi n$. In what follows,
we will for simplicity restrict ourselves to the monopoles
possessing the minimal charge only, i.e. set $n=1$, although the generalization to an
arbitrary $n$ is straightforward.
Note that the origin of root vectors in eq.~(\ref{et6}) is the fact that
monopole charges are distributed along them. Further,
$\Phi_i=\left|\Phi_i\right|{\rm e}^{i\theta_i}$ are the
dual Higgs fields, which describe the condensates of monopoles, and
${\bf F}_{\mu\nu}=\partial_\mu{\bf B}_\nu-\partial_\nu{\bf B}_\mu$ is the
field-strength tensor of the
$(N-1)$-component ``magnetic'' potential ${\bf B}_\mu$. The latter is dual
to the ``electric'' potential, whose components are diagonal gluons.
Since the SU($N$)-group is special, the phases $\theta_i$ of the
dual Higgs fields are related to each other by the constraint
$\sum\limits_{i}^{}\theta_i=0$, which is imposed by introducing
the corresponding $\delta$-function into the r.h.s. of eq.~(\ref{et6}).
Next, $\tilde {\cal O}_{\mu\nu}\equiv\frac12\varepsilon_{\mu\nu\lambda\rho}
{\cal O}_{\lambda\rho}$, and
${\bf F}_{\mu\nu}^k$ is the field-strength tensor
of $k$ test quarks with colors $a_i$, which move along a 
contour $C$. This tensor obeys the equation
$\partial_\mu\tilde {\bf F}_{\mu\nu}^k=g{\bf M}_k j_\nu$,
where ${\bf M}_k\equiv\sum\limits_{i=1}^{k}{\bf m}_{a_i}$, 
$j_\mu(x)=\oint\limits_{C}^{}dx_\mu(\tau)\delta(x-x(\tau))$,
${\bf m}_{a_i}$ is a weight vector of the group SU($N$), and $a_i$ may take values 
$1,\ldots, N$.
Thus ${\bf F}_{\mu\nu}^k=-g{\bf M}_k\tilde\Sigma_{\mu\nu}$,
where $\Sigma_{\mu\nu}(x)=\int\limits_{\Sigma}^{}d\sigma_{\mu\nu}
(x(\xi))\delta(x-x(\xi))$ is the vorticity tensor current associated
with the world-sheet $\Sigma$ of
the open electric string, bounded by the contour $C$.
From now on, we will omit the normalization constant in front of
all the functional integrals, implying for any $k$ the normalization condition
${\cal Z}_k\left[C=0\right]=1$.

Next, the phases of the dual Higgs fields can be decomposed into
multivalued and single-valued (also called singular and
regular, respectively) parts, $\theta_i= \theta_i^{\rm sing}+
\theta_i^{\rm reg}$. The fields $\theta_i^{\rm sing}$ describing
closed dual strings are related to the world-sheets $\Sigma_i$ of
these strings by means of the equation

\begin{equation}
\label{suz3}
\varepsilon_{\mu\nu\lambda\rho}\partial_\lambda\partial_\rho
\theta_i^{\rm sing}(x)=2\pi\Sigma_{\mu\nu}^i(x)\equiv
2\pi\int\limits_{\Sigma_i}^{}d\sigma_{\mu\nu}\left(x^{(i)}(\xi)\right)
\delta\left(x-x^{(i)}(\xi)\right).
\end{equation}
This equation is the covariant formulation of the 4-d analogue of the
Stokes' theorem for $\partial_\mu\theta_i$, written in the local
form. In eq.~(\ref{suz3}), $x^{(i)}(\xi)\equiv x_\mu^{(i)}(\xi)$ is a
vector, that parameterizes the world-sheet $\Sigma_i$ with $\xi=(\xi^1,
\xi^2)$ standing for the 2-d coordinate. As far as the regular parts
of the phases, $\theta_i^{\rm reg}$, are concerned, these describe
single-valued fluctuations around closed strings, which are described
by $\theta_i^{\rm sing}$.  Note that, owing to the one-to-one
correspondence between $\theta_i^{\rm sing}$ and $\Sigma_i$,
established by eq.~(\ref{suz3}), the integration over $\theta_i^{\rm
sing}$ is implied in the sense of a certain prescription of the
summation over world-sheets of closed strings. One of such
prescriptions, corresponding to the dilute plasma of closed strings,
will be considered in Section~5. Further, by virtue of
eq.~(\ref{suz3}), it is also possible to demonstrate that the
integration measure ${\cal D}\theta_i$ becomes factorized into the
product ${\cal D}\theta_i^{\rm sing}{\cal D}\theta_i^{\rm reg}$.
Apparently, the constraint imposed by the $\delta$-function
$\delta\left(\sum\limits_{i}^{}\theta_i\right)$ becomes also split as
$\delta\left(\sum\limits_{i}^{}\theta_i^{\rm
sing}\right)\delta\left(\sum\limits_{i}^{}\theta_i^{\rm reg}\right)$.

Let us further expand $|\Phi_i|$ in eq.~(\ref{et6}) as
$|\Phi_i|=\eta+\frac{\varphi_i}{\sqrt{2}}$ and perform the gauge
transformation $g_m{\bf q}_i{\bf B}_\mu^{\rm new}=g_m{\bf q}_i{\bf
B}_\mu-\partial_\mu\theta_i$, noticing that, according to
eq.~(\ref{suz3}),
$(\partial_\mu\partial_\nu-\partial_\nu\partial_\mu)\theta_i^{\rm
sing}= 2\pi\tilde\Sigma_{\mu\nu}^i$. The constraint
$\sum\limits_{i}^{}\theta_i^{\rm sing}=0$ leads to the constraint for
world-sheets of closed strings, which can be imposed by the
$\delta$-function $\delta\left(\sum\limits_{i}^{}
\Sigma_{\mu\nu}^i\right)$. Instead, the constraint imposed by
$\delta\left(\sum\limits_{i}^{}\theta_i^{\rm reg}\right)$ can be
lifted to the exponent upon the introduction of some Lagrange
multiplier. By virtue of the fact that $\sum\limits_{i}^{}{\bf
q}_i=0$, the integration over this multiplier can be shown~\cite{su3a}
to eventually result in an inessential overall constant factor.  If we
further use the orthonormality of roots,

\begin{equation}
\label{ortho}
\sum\limits_{i}^{}q_i^\alpha q_i^\beta=\frac{N}{2}\delta^{\alpha\beta},
\end{equation} 
where $\alpha, \beta=1,\ldots,
N-1$, we obtain the following Lagrangian 

$$
{\cal L}_k=
\frac{1}{2Ng_m^2}\sum\limits_{i}^{}\left[{\bf q}_i\left(g_m{\bf F}_{\mu\nu}-4\pi{\bf M}_k\tilde\Sigma_{\mu\nu}\right)+
2\pi\tilde\Sigma_{\mu\nu}^i\right]^2+\frac{m^2}{2}{\bf B}_\mu^2+$$

\begin{equation}
\label{Lk}
+\sum\limits_{i}^{}\left[\frac12(\partial_\mu\varphi_i)^2+\frac{m_H^2}{2}\varphi_i^2+
\sqrt{2}g_m^2\eta\varphi_i({\bf q}_i{\bf B}_\mu)^2\right].
\end{equation}
The masses of the dual vector boson and the dual Higgs field here read $m=g_m\eta\sqrt{N}$ and $m_H=2\eta\sqrt{\lambda}$.

\section{London limit}

In terms of the Landau--Ginzburg parameter
$\kappa\equiv\frac{m_H}{m}$, the London limit (LL) is defined as
$\ln\kappa\gg 1$.  Let us now proceed with the string representation
of the model~(\ref{Lk}) in this limit.  Notice that, since we would
like our model to be consistent with QCD, we must have
$g=\sqrt{\bar\lambda/N}$, where $\bar\lambda$ remains finite in the
large-$N$ limit. The definition of the London limit then yields

\begin{equation}
\label{lam}
\lambda\gg\frac{(2\pi {\rm e}N)^2}{\bar\lambda}.
\end{equation} 
In general, in order to have confinement (i.e. the type-II dual
superconductivity) in the large-$N$ limit, one should demand that
$\lambda$ grows with $N$ at least as ${\cal O}(N^2)$. If $\lambda$
grows with $N$ faster than $N^2$, $\lambda={\cal O}(N^{2+\epsilon})$
where $\epsilon>0$, $\kappa$ grows with $N$ too, $\kappa={\cal
O}(N^{\epsilon/2})$, making the London limit deeper.  In what follows,
we will adopt the minimal requirement, $\lambda={\cal O}(N^2)$,
necessary for confinement in the large-$N$ limit of the
model~(\ref{et6}). As we have seen, only in this case, $\kappa$ is
$N$-independent.

In the London limit, the partition function of our model has the form

\begin{equation}
\label{partfunct}
{\cal Z}_k^{\rm LL}=\int\left(\prod\limits_{i}^{}\left[d\Sigma_{\mu\nu}^i\right]\right)
\delta\left(\sum\limits_{i}^{}
\Sigma_{\mu\nu}^i\right)\int{\cal D}{\bf B}_\mu\exp\left(-\int d^4x{\cal L}_k^{\rm LL}\right),
\end{equation}
where the Lagrangian reads

$$ {\cal L}_k^{\rm LL}=\frac{1}{2Ng_m^2}\sum\limits_{i}^{} \left[{\bf
q}_i\left(g_m{\bf F}_{\mu\nu}-4\pi{\bf
M}_k\tilde\Sigma_{\mu\nu}\right)+
2\pi\tilde\Sigma_{\mu\nu}^i\right]^2+\frac{m^2}{2}{\bf B}_\mu^2.$$ The
symbol $\left[d\Sigma_{\mu\nu}^i\right]$ in eq.~(\ref{partfunct}) is a
formal expression for the sum over string world-sheets, whose concrete
form will be specified below. 

A natural way to satisfy the constraint
$\sum\limits_{i}^{}\Sigma_{\mu\nu}^i=0$ is to set
$\Sigma_{\mu\nu}^i={\bf q}_i{\bf S}_{\mu\nu}$, where ${\bf
S}_{\mu\nu}$ are then no longer subject to any constraint. The
Lagrangian then takes the form

$$
{\cal L}_k^{\rm LL}=\frac{1}{4g_m^2}\left(g_m{\bf F}_{\mu\nu}-4\pi{\bf M}_k\tilde\Sigma_{\mu\nu}+
2\pi\tilde{\bf S}_{\mu\nu}\right)^2+\frac{m^2}{2}{\bf B}_\mu^2.$$
To perform the integration over ${\bf B}_\mu$, let us linearize the squares in this Lagrangian upon the 
introduction of two auxiliary fields, ${\bf h}_{\mu\nu}$ and ${\bf k}_\mu$, as follows:

\begin{equation}
\label{k}
{\cal L}_k^{\rm LL}=\frac14{\bf h}_{\mu\nu}^2+\frac{i}{2}\tilde{\bf h}_{\mu\nu}
\left({\bf F}_{\mu\nu}-g{\bf M}_k\tilde\Sigma_{\mu\nu}+
\frac{g}{2}\tilde{\bf S}_{\mu\nu}\right)+{\bf k}_\mu^2+i\sqrt{2}m{\bf k}_\mu{\bf B}_\mu.
\end{equation}
The integration over ${\bf B}_\mu$ then yields 

\begin{equation}
\label{ll1}
{\bf k}_\mu=\frac{1}{m\sqrt{2}}\partial_\nu\tilde{\bf h}_{\nu\mu},
\end{equation}
and we obtain

\begin{equation}
\label{ll2}
{\cal L}_k^{\rm LL}=\frac{1}{2m^2}(\partial_\mu\tilde{\bf h}_{\mu\nu})^2+\frac14
{\bf h}_{\mu\nu}^2+\frac{ig}{2}{\bf h}_{\mu\nu}\left(\frac{{\bf S}_{\mu\nu}}{2}-{\bf M}_k\Sigma_{\mu\nu}\right).
\end{equation}
To perform the integration over the so-called Kalb-Ramond field ${\bf h}_{\mu\nu}$, 
notice that the general solution to eq.~(\ref{ll1}) with respect to 
${\bf h}_{\mu\nu}(x)$ reads

$$ {\bf
h}_{\mu\nu}(x)=-\sqrt{2}m\varepsilon_{\mu\nu\lambda\rho}\partial_\lambda^x\int
d^4x' D_0(x-x'){\bf k}_\rho(x')+\partial_\mu{\bf
C}_\nu(x)-\partial_\nu{\bf C}_\mu(x),$$ where ${\bf C}_\mu$ is an
arbitrary vector field, and $D_0(x)=1/(4\pi^2x^2)$ is the Coulomb
propagator.  To fix ${\bf C}_\mu=0$ is equivalent to impose the
constraint $\partial_\mu{\bf h}_{\mu\nu}=0$.  On the other hand,
because of the coupling of ${\bf h}_{\mu\nu}$ to the open-string world-sheet,
$-\frac{ig}{2}{\bf M}_k{\bf h}_{\mu\nu}\Sigma_{\mu\nu}$, the
saddle-point value of ${\bf h}_{\mu\nu}$, which saturates the
respective Gaussian integral, does not obey the constraint
$\partial_\mu{\bf h}_{\mu\nu}=0$.  To make the integration consistent,
we should therefore promote ${\bf h}_{\mu\nu}$ by making ${\bf C}_\mu$
non-vanishing. For such ${\bf h}_{\mu\nu}$, $(\partial_\mu\tilde{\bf
h}_{\mu\nu})^2$ in eq.~(\ref{ll2}) can be replaced by $\frac16{\bf
H}_{\mu\nu\lambda}^2$, where ${\bf H}_{\mu\nu\lambda}=\partial_\mu{\bf
h}_{\nu\lambda}+ \partial_\lambda{\bf h}_{\mu\nu}+\partial_\nu{\bf
h}_{\lambda\mu}$ is the field-strength tensor of ${\bf h}_{\mu\nu}$.
The integration over ${\bf h}_{\mu\nu}$ in the resulting theory with
the Lagrangian

\begin{equation}
\label{H}
{\cal L}_k^{\rm LL}=\frac{1}{12m^2}{\bf H}_{\mu\nu\lambda}^2+\frac14
{\bf h}_{\mu\nu}^2+\frac{ig}{2}{\bf h}_{\mu\nu}\left(\frac{{\bf
S}_{\mu\nu}}{2}-{\bf M}_k\Sigma_{\mu\nu}\right)
\end{equation}
is straightforward (see e.g.~\cite{epj} for details) and yields 

$$
{\cal Z}_k^{\rm LL}=\exp\left\{-\frac{(g{\bf M}_k)^2}{2}
\left[\oint\limits_{C}^{}dx_\mu
\oint\limits_{C}^{}dx'_\mu D_m(x-x')+\right.\right.$$

$$\left.\left.+\frac{m^2}{2}\int
d^4xd^4x'\Sigma_{\mu\nu}(x)D_m(x-x')\Sigma_{\mu\nu}(x')\right]\right\}\times$$

$$ \times\int\left[d{\bf S}_{\mu\nu}\right]
\exp\left[-\left(\frac{gm}{4}\right)^2\int d^4xd^4x'{\bf
S}_{\mu\nu}(x)D_m(x-x'){\bf S}_{\mu\nu}(x')+\right.$$

\begin{equation}
\label{ll3}
\left.+\left(\frac{gm}{2}\right)^2{\bf M}_k\int d^4xd^4x'{\bf
S}_{\mu\nu}(x)D_m(x-x')\Sigma_{\mu\nu}(x')\right],
\end{equation}
where $D_m=mK_1(m|x|)/(4\pi^2|x|)$ is the Yukawa propagator. The
formal measure $\left[d{\bf S}_{\mu\nu}\right]$ here, which replaces
the measure $\left[d\Sigma_{\mu\nu}^i\right]$ in
eq.~(\ref{partfunct}), implies a certain prescription of the summation
over world sheets of closed strings. For the dilute-plasma model of
closed strings, this measure will be found in Section~5.

According to the first term on the r.h.s. of eq.~(\ref{ll3}), the
Yukawa part of the potential satisfies the Casimir-scaling law,
since ${\bf M}_k^2\equiv C_k=\frac{k(N-k)}{2N}$~\cite{jh}. Further, in
the leading semiclassical approximation we are considering in this
and next sections, the integral over small-sized closed strings can be
disregarded, and the remaining
$\Sigma_{\mu\nu}\times\Sigma_{\mu\nu}$-interaction produces the
Casimir scaling also for the confining part of the potential. In fact,
extracting from this interaction the string tension according to the
respective general formula~\cite{mpla} (cf. also~\cite{plbN}), we
obtain $\sigma_k=C_k\bar\sigma$, where $\bar\sigma=4\pi
N\eta^2\ln\kappa$. Note that, for $\sigma_1$ to be $N$-independent
as the quark--antiquark string tension in QCD, we should have

\begin{equation}
\label{eta}
\eta\sim\frac{1}{\sqrt{N\ln\kappa}}={\cal O}(N^{-1/2}).
\end{equation}
In what follows, we will address corrections to the Casimir scaling,
that appear from the deviation from the London limit, as well as
corrections produced by closed dual strings.

\section{Corrections due to the deviation from the London limit}

Let us consider the Lagrangian~(\ref{Lk}) with $\Sigma_{\mu\nu}^i=0$.
Introducing the ${\bf h}_{\mu\nu}$-field in the same way as in the London limit, 
we obtain the following expression:

$$ {\cal L}_k=\frac14{\bf
h}_{\mu\nu}^2+\frac12(\partial_\mu\varphi_i)^2+\frac{m_H^2}{2}
\varphi_i^2-\frac{ig}{2}{\bf M}_k{\bf h}_{\mu\nu}\Sigma_{\mu\nu}+i{\bf
B}_\mu\partial_\nu\tilde{\bf h}_{\mu\nu}
+\frac{m^2}{2}B_\mu^\alpha\left(\delta^{\alpha\beta}+\xi^{\alpha\beta}\right)B_\mu^\beta,$$
where the tensor $\xi^{\alpha\beta}\equiv\frac{2\sqrt{2}}{\eta
N}\sum\limits_{i}^{}\varphi_i q_i^\alpha q_i^\beta$ is apparently
symmetric. Performing the Gaussian integration over ${\bf B}_\mu$, we
arrive at the following substitution:

\begin{equation}
\label{gauss}
i{\bf B}_\mu\partial_\nu\tilde{\bf h}_{\mu\nu}
+\frac{m^2}{2}B_\mu^\alpha\left(\delta^{\alpha\beta}+\xi^{\alpha\beta}\right)B_\mu^\beta
\longrightarrow \frac{1}{2m^2}(\partial_\nu\tilde
h_{\mu\nu}^\alpha)\left(\delta^{\alpha\beta}-\xi^{\alpha\beta}\right)
(\partial_\lambda\tilde h_{\mu\lambda}^\beta).
\end{equation}
It can be shown (see Appendix~A for details) that
$\det{}^{-1/2}\left[\hat 1+\Xi\right]$, where $\hat 1$ and $\Xi$ are
the unit and the $\xi^{\alpha\beta}$-matrices, produces a
renormalization of $m$ and $m_H$, which does not violate the
London limit condition, $\ln\kappa\gg 1$. Notice also that we have
retained only the term linear in $\Xi$ on the r.h.s. of
eq.~(\ref{gauss}). As we will see below, this linear term eventually
produces a correction to $\sigma_k$, which we are looking for. Instead, the
omitted $\Xi^2$-term is shown in Appendix~A to produce merely
an inessential correction to $m_H$, which is smaller than $m_H$ in the
factor ${\cal O}\left(\frac{1}{\kappa\sqrt{N}}\right)$.

Next, the $\delta^{\alpha\beta}$-term on the r.h.s. of
eq.~(\ref{gauss}) is clearly the kinetic term of the Kalb--Ramond
field, that is the first term on the r.h.s. of
eq.~(\ref{ll2}). Instead, the $\xi^{\alpha\beta}$-term on the
r.h.s. of eq.~(\ref{gauss}) is the Higgs-inspired correction, which we
will denote as $C^{\alpha\beta}\xi^{\alpha\beta}$, where
$C^{\alpha\beta}\equiv-\frac{1}{2m^2}(\partial_\nu\tilde
h_{\mu\nu}^\alpha) (\partial_\lambda\tilde h_{\mu\lambda}^\beta)$. The
Gaussian integration over $\varphi_i$ then leads to the following
substitution:

\begin{equation}
\label{co}
\frac12(\partial_\mu\varphi_i)^2+\frac{m_H^2}{2}\varphi_i^2+\xi^{\alpha\beta}C^{\alpha\beta}\longrightarrow
-\frac{4}{(\eta N)^2}\sum\limits_{i}^{}q_i^\alpha q_i^\beta q_i^\gamma q_i^\delta C^{\alpha\beta}(x)
\int d^4x' D_{m_H}(x-x')C^{\gamma\delta}(x').
\end{equation}
Analogously to eq.~(\ref{ortho}), 
the orthonormality of roots yields the following formula (cf. Ref.~\cite{jh}):

$$
\sum\limits_{i}^{} q_i^\alpha q_i^\beta q_i^\gamma q_i^\delta=\frac{N}{2(N+1)}
\left(\delta^{\alpha\beta}\delta^{\gamma\delta}+\delta^{\alpha\gamma}\delta^{\beta\delta}+
\delta^{\alpha\delta}\delta^{\beta\gamma}\right).$$
Using this, we finally obtain the following Lagrangian:

$$
{\cal L}_k=-C^{\alpha\alpha}+\frac14{\bf h}_{\mu\nu}^2
-\frac{ig}{2}{\bf M}_k{\bf h}_{\mu\nu}\Sigma_{\mu\nu}-
$$

\begin{equation}
\label{corr}
-\frac{2}{\eta^2N(N+1)}\int d^4x' D_{m_H}(x-x')\left[C^{\alpha\alpha}(x)C^{\beta\beta}(x')+
2C^{\alpha\beta}(x)C^{\alpha\beta}(x')\right].
\end{equation}
To proceed, notice that, according to eq.~(\ref{k}), the
physical meaning of the field ${\bf k}_\mu$ is the monopole current,
which couples to the dual gauge field ${\bf B}_\mu$ and also possesses
its own self-interaction terms. (Since for the massive vector field
one always has $\partial_\mu{\bf B}_\mu=0$, this current is
automatically conserved.) The monopole current is defined in terms of
the Kalb--Ramond field by means of eq.~(\ref{ll1}). Using this
correspondence, it is straightforward to reformulate the action,
corresponding to the obtained Lagrangian~(\ref{corr}), in terms of
monopole currents:

$$
\int d^4x {\cal L}_k=\int d^4x{\bf k}_\mu^2+
m^2\int d^4x d^4x' {\bf k}_\mu(x)D_0(x-x'){\bf k}_\mu(x')+$$

$$
+igm\sqrt{2}{\bf M}_k
\int d^4x d^4x' \tilde\Sigma_{\mu\nu}(x){\bf k}_\nu(x')\partial_\mu^xD_0(x-x')-
\frac{2}{\eta^2N(N+1)}\int d^4xd^4x' D_{m_H}(x-x')\times$$

\begin{equation}
\label{higgs}
\times\left[k_\mu^\alpha(x)k_\mu^\alpha(x)
k_\nu^\beta(x')k_\nu^\beta(x')+2k_\mu^\alpha(x)k_\mu^\beta(x)
k_\nu^\alpha(x')k_\nu^\beta(x')\right],
\end{equation}
where the terms are presented in the same order as they stand in
eq.~(\ref{corr}).  The first and second terms on the r.h.s. of this
equation are clearly the mass term of the current and the Coulomb
interaction between currents, respectively. The third term has a
form similar to the Gauss linking number of a closed surface $\Sigma$
and a closed contour $\Gamma$, $L(\Sigma,\Gamma)=\int d^4xd^4x'
\tilde\Sigma_{\mu\nu}(x)j_\nu(x')\partial_\mu^xD_0(x-x')$, where
$j_\mu(x)\equiv\oint\limits_{\Gamma}^{}dx_\mu(\tau)
\delta(x-x(\tau))$. Our formula differs from the Gauss one in two
respects: first, the surface is open, and second, instead of the
classical current $j_\mu$ of a point-like particle, localized along a
closed trajectory, we have the quantum field, $gm\sqrt{2}{\bf M}_k{\bf
k}_\mu$, distributed over the whole space-time.

Let us now estimate the Higgs-inspired correction, given by the last term on the r.h.s. of eq.~(\ref{higgs}).
Notice that, according to eq.~(\ref{eta}), $m$ depends on $N$ as
$m=g_m\eta\sqrt{N}\sim\sqrt{N}\cdot\frac{1}{\sqrt{N}}\cdot\sqrt{N}=\sqrt{N}$, and $m_H=\kappa m$, where
$\kappa$ has been naturally chosen $N$-independent 
(cf. the first paragraph of Section~3). Therefore, $m_H$ grows with
$N$ as

\begin{equation}
\label{MH}
m_H={\cal O}(N^{1/2}).
\end{equation}
The heaviness of the Higgs bosons, implied in both London and large-$N$ limits,
enables us to write, in the leading $1/m_H$-approximation,
the following expression for the correction under study:

\begin{equation}
\label{leadord}
-\frac{2}{(\eta m_H)^2N(N+1)}\int d^4x\left[({\bf k}_\mu^2)^2+2({\bf k}_\mu{\bf k}_\nu)^2\right].
\end{equation}
After the variation of the total action with respect to
$k_\mu^\alpha$, the color structure of the saddle-point
equation thus obtained prescribes to seek ${\bf k}_\mu$ in the form ${\bf
M}k_\mu$. The equation for $k_\mu$ then reads

$$
\left(1-\frac{12C_k}{(\eta m_H)^2N(N+1)}k_\rho^2\right)k_\mu+m^2\int d^4x'D_0(x-x')k_\mu(x')=$$

$$
=-\frac{igm}{\sqrt{2}}\int d^4x' \tilde\Sigma_{\mu\nu}(x')\partial_\nu^xD_0(x-x').
$$ 
The leading-order part of this equation, without the
$k_\rho^2$-term, can easily be converted into the differential form
by acting with $\partial^2$ onto both its sides. The resulting
equation
$(\partial^2-m^2)k_\mu=\frac{igm}{\sqrt{2}}\partial_\nu\tilde\Sigma_{\mu\nu}$
leads to the following leading-order saddle-point expression for
$k_\mu$: $k_\mu(x)=-\frac{igm}{\sqrt{2}}\int d^4x'
D_m(x-x')\partial_\nu\tilde\Sigma_{\mu\nu}(x')$, which should further
be substituted into eq.~(\ref{leadord}), 

\begin{equation}
\label{corr1}
-\frac{6(C_k)^2}{(\eta m_H)^2N(N+1)}\int d^4x(k_\mu^2)^2.
\end{equation}
Among all the terms contained here in $k_\mu^2$, only the surface$\times$surface one,

$$ -\frac{(gm)^2}{4}\int
d^4x_1d^4x_2\Sigma_{\mu\nu}(x_1)\Sigma_{\mu\nu}(x_2)\partial_\alpha^{x_1}D_m(x-x_1)
\partial_\alpha^{x_2}D_m(x-x_2),
$$ produces the desired correction to the string tension. Indeed, the
integral structure of the respective part of the
correction~(\ref{corr1}),

\begin{equation}
\label{correc}
-\frac{3(C_k)^2(gm)^4}{8N(N+1)(\eta m_H)^2}\int d\sigma_{\mu\nu}(x_1)d\sigma_{\mu\nu}(x_2)
d\sigma_{\lambda\rho}(x_3)d\sigma_{\lambda\rho}(x_4)\cdot J,
\end{equation}
where $J\equiv \partial_\alpha^{x_1}\partial_\alpha^{x_2}
\partial_\beta^{x_3}\partial_\beta^{x_4}\int
d^4x\prod\limits_{l=1}^{4}D_m(x-x_l)$, is the same as appears in the
3-d SU($N$) Georgi--Glashow model due to the non-diluteness of the
monopole plasma~\cite{jh}. In that paper, it was shown that this
structure does produce a correction to the string tension.  
It can be shown that, similarly to the 3-d SU($N$) Georgi--Glashow
model, this correction behaves with $N$ as $(C_k)^2/N$.

With the account for the obtained correction, the $N$-ality dependence
of the ratio of string tensions is given by the formula

$$
\frac{\sigma_k}{\sigma_1}=\frac{k(N-k)}{N-1}\left[1+\alpha(N)\frac{(k-1)(N-k-1)}{N}\right],$$
where the coefficient $\alpha(N)\sim N^{-1}$ and therefore, at 
fixed $k$, the whole correction vanishes in the large-$N$ limit.  As
well as the Casimir-scaling term, the obtained leading Higgs-inspired
correction is apparently invariant under the interchange of quarks and
antiquarks, $k\leftrightarrow (N-k)$.

\section{Corrections due to closed strings}

To study the grand canonical ensemble of closed strings, it is
necessary to replace ${\bf S}_{\mu\nu}$ in eq.~(\ref{H}) (with
$\Sigma_{\mu\nu}$ set for a while equal to zero) by
$\sum\limits_{a}^{}n_a{\bf S}_{\mu\nu}^a$, where the $n_a$ stand for
winding numbers. It is known~\cite{22, moresu3} that one may restrict
oneself to closed strings possessing the minimal winding numbers,
$n_a=\pm 1$. That is merely because the energy of a single closed
string is a quadratic function of its flux, owing to which it is
energetically favorable for the vacuum to maintain two closed strings
of a unit flux, rather than one string of the double flux.

Then, taking into account that the plasma of closed strings is dilute,
one can perform the summation over the grand canonical ensemble 
of these objects, which modifies the Lagrangian~(\ref{H}), 
with $\Sigma_{\mu\nu}=0$, as follows:

\begin{equation}
\label{lgc}
{\cal L}_{\rm gr.{\,}can.}=\frac{1}{12m^2}{\bf H}_{\mu\nu\lambda}^2+\frac14
{\bf h}_{\mu\nu}^2-2\zeta\cos\left(\frac{g}{4}\frac{|{\bf h}_{\mu\nu}|}{\Lambda^2}\right).
\end{equation}
Here $\Lambda\equiv\sqrt{L/a^3}$ is a UV momentum cut-off with $L$ and
$a$ denoting the characteristic distances between closed strings and
their typical sizes, respectively.  In the dilute-plasma approximation
under study, $a\ll L$ and $\Lambda\gg a^{-1}$. Next, $\zeta\propto
{\rm e}^{-S_0}$ stands for the fugacity (Boltzmann factor) of a single
string, which has the dimension $({\rm mass})^4$, with $S_0$ denoting
the action of a single string, $S_0\sim\sigma_1 a^2$. Finally, it is
assumed that closed strings are not too small, namely $a\ge{\cal
O}\left(\frac{1}{gm}\right)$, so that $S_0\gg 1$, and the mean density
of the plasma, $2\zeta$, is exponentially small, i.e. the plasma is
dilute.

Note also that, because of the Debye screening of the dual vector boson in
the plasma of closed strings, its mass increases. This is clearly seen
from eq.~(\ref{lgc}), by the increase of the mass of the Kalb--Ramond
field, which represents this boson:

\begin{equation}
\label{M}
m^2\longrightarrow M^2=m^2\left(1+\frac{g^2\zeta}{4\Lambda^4}\right).
\end{equation}

To study corrections to the $k$-string tension produced by closed
strings, we will need to know correlation functions of these strings
in the plasma. To obtain an expression for the generating functional
of such correlation functions, one needs the theory to be formulated in
terms of dynamical vorticity tensor currents. This can be done by
recalling that, for closed strings:

$$\exp\left\{-\int d^4x\left[\frac{1}{12m^2}{\bf H}_{\mu\nu\lambda}^2+\frac14
{\bf h}_{\mu\nu}^2\right]\right\}=$$

$$=\int {\cal D}{\bf S}_{\mu\nu}\exp\left[-\left(\frac{gm}{4}\right)^2
\int d^4xd^4x'{\bf S}_{\mu\nu}(x)D_m(x-x'){\bf
S}_{\mu\nu}(x')-\frac{ig}{4}\int d^4x{\bf h}_{\mu\nu} {\bf
S}_{\mu\nu}\right].$$ 
The Kalb--Ramond field can then be integrated out by solving the
saddle-point equation stemming from the respective part of the
Lagrangian,

$$ -2\zeta\cos\left(\frac{g}{4}\frac{|{\bf
    h}_{\mu\nu}|}{\Lambda^2}\right)+\frac{ig}{4}{\bf h}_{\mu\nu} {\bf
  S}_{\mu\nu}.
$$ 
This yields the following expression for the partition function of
the grand canonical ensemble of closed strings in terms of their
vorticity tensor currents:

\begin{equation}
\label{zgc}
{\cal Z}_{\rm gr.{\,}can.}=\int {\cal D}{\bf S}_{\mu\nu}\exp\left\{-\left[\left(\frac{gm}{4}\right)^2
\int d^4xd^4x'{\bf S}_{\mu\nu}(x)D_m(x-x'){\bf S}_{\mu\nu}(x')+V[{\bf S}_{\mu\nu}]\right]\right\},
\end{equation}
where the potential $V[{\bf S}_{\mu\nu}]$ reads

\begin{equation}
\label{v}
V[{\bf S}_{\mu\nu}]=\int d^4x\left\{\Lambda^2|{\bf S}_{\mu\nu}|\ln\left[
\frac{\Lambda^2}{2\zeta}|{\bf S}_{\mu\nu}|+\sqrt{1+\left(
\frac{\Lambda^2}{2\zeta}|{\bf S}_{\mu\nu}|\right)^2}\right]-2\zeta
\sqrt{1+\left(
\frac{\Lambda^2}{2\zeta}|{\bf S}_{\mu\nu}|\right)^2}\right\}.
\end{equation}
(Note that the operator, which describes the density of plasma at the
point $x$, is $\Lambda^2|{\bf S}_{\mu\nu}(x)|$.)  Comparing now
eqs.~(\ref{zgc}), (\ref{v}) with eq.~(\ref{ll3}), we see that, in the
dilute-plasma model of closed strings, the formal measure $[d{\bf
S}_{\mu\nu}]$ concretizes as ${\cal D}{\bf S}_{\mu\nu}{\rm e}^{-V[{\bf
S}_{\mu\nu}]}$.

Corrections to the string tension $\sigma_k$ of the open world-sheet $\Sigma$ stem from the last
term in the following expression for the partition function~(\ref{ll3}):

$$
-\ln {\cal Z}_k^{\rm LL}=$$

$$=\frac{(g{\bf M}_k)^2}{2}\left[\oint\limits_{C}^{}dx_\mu
\oint\limits_{C}^{}dx'_\mu D_m(x-x')+\frac{m^2}{2}
\int d^4xd^4x'\Sigma_{\mu\nu}(x)D_m(x-x')\Sigma_{\mu\nu}(x')\right]-$$

$$
-\ln\left<\exp\left[\left(\frac{gm}{2}\right)^2{\bf M}_k
\int d^4xd^4x'{\bf S}_{\mu\nu}(x)
D_m(x-x')\Sigma_{\mu\nu}(x')\right]\right>,$$
where the average $\left<\ldots\right>$ over closed strings is now defined according to eqs.~(\ref{zgc}),~(\ref{v}).
By virtue of the cumulant expansion, this term can be written as

$$
-\sum\limits_{n=1}^{\infty}\left(\frac{gm}{2}\right)^{2n}M_k^{\alpha_1}\cdots M_k^{\alpha_n}\times$$

$$\times
\int d^4x_1\cdots d^4x_n\sigma_{\mu_1\nu_1}(x_1)\cdots\sigma_{\mu_n\nu_n}(x_n)\left<\left<
S_{\mu_1\nu_1}^{\alpha_1}(x_1)\cdots S_{\mu_n\nu_n}^{\alpha_n}(x_n)\right>\right>.$$
Here, $\sigma_{\mu\nu}(x)\equiv\int d^4x'D_m(x-x')\Sigma_{\mu\nu}(x')$, and $\left<\left<\cdots\right>\right>$
denotes a one-particle irreducible average (cumulant) of closed strings.
Since the action corresponding to the partition function~(\ref{zgc}) contains only powers of ${\bf S}_{\mu\nu}(x)
{\bf S}_{\mu\nu}(x')$, cumulants of odd orders vanish, whereas a cumulant of an even order $n$ has the form

$$\left<\left<
S_{\mu_1\nu_1}^{\alpha_1}(x_1)\cdots S_{\mu_n\nu_n}^{\alpha_n}(x_n)\right>\right>=$$

$$=\delta^{\alpha_1\alpha_2}
\cdots\delta^{\alpha_{n-1}\alpha_n}f_{\mu_1\nu_1,\mu_2\nu_2}(x_1-x_2)\cdots
f_{\mu_{n-1}\nu_{n-1},\mu_n\nu_n}(x_{n-1}-x_n)+ {\rm permutations}.$$
Here,
$f_{\mu_1\nu_1,\mu_2\nu_2}(x_1-x_2)=\varepsilon_{\mu_1\nu_1\lambda_1\rho}\varepsilon_{\mu_2\nu_2\lambda_2\rho}
\partial_{\lambda_1}^{x_1}\partial_{\lambda_2}^{x_2}{\cal D}(x_1-x_2)$
with ${\cal D}$ standing for a function whose concrete form
for the case of a very dilute plasma will be made clear in a
moment. As for the Lorentz structure of the $f$-tensors, it stems from
the condition of closeness of strings, $\partial_\mu{\bf
S}_{\mu\nu}=0$.  The color structure of the cumulant produces, for
$n\equiv 2l$, the factor $({\bf M}_k^2)^l=(C_k)^l$. The $N$-ality
dependence of $\sigma_k$ is therefore defined by the following
formula:

\begin{equation}
\label{tens}
\sigma_k=C_k\bar\sigma+\sum\limits_{l=1}^{\infty}\sigma^{(l)}(C_k)^l,
\end{equation} 
where $\sigma^{(l)}$ are ($N$-dependent) coefficients of dimension
$[{\rm mass}]^2$. Note that the term with $l=2$ in this equation
produces a correction to the string tension, which has the same
$(C_k)^2$-dependence as the leading Higgs-inspired correction found in
the previous section.

As an example, let us finally present the lowest non-trivial two-point
correlation function of closed strings, which can be derived in the
approximation when the plasma is very dilute, i.e. its density is even
lower than the (already exponentially small) mean one, $2\zeta$.  In
that case, the potential~(\ref{v}) becomes a quadratic functional.
Including the source term, $\int d^4x{\bf J}_{\mu\nu}{\bf
S}_{\mu\nu}$, into the square brackets on the r.h.s. of
eq.~(\ref{zgc}), we obtain for this Gaussian integral:

$${\cal Z}_{\rm gr.{\,}can.}[{\bf J}_{\mu\nu}]\simeq\frac{1}{{\cal Z}_{\rm gr.{\,}can.}[0]}
\int {\cal D}{\bf S}_{\mu\nu}\exp\left\{-\left[\left(\frac{gm}{4}\right)^2
\int d^4xd^4x'{\bf S}_{\mu\nu}(x)D_m(x-x'){\bf S}_{\mu\nu}(x')+\right.\right.$$

$$
\left.\left.+\int d^4x\left(-2\zeta+
\frac{\Lambda^4}{4\zeta}{\bf S}_{\mu\nu}^2
+{\bf J}_{\mu\nu}{\bf S}_{\mu\nu}\right)\right]\right\}
=\exp\left[-\int d^4x d^4x'{\bf J}_{\mu\nu}(x){\cal G}(x-y){\bf J}_{\mu\nu}(x')
\right],$$
where ${\cal G}(x)\equiv\frac{\zeta}{\Lambda^4}(\partial^2-m^2)D_M(x)$, and the mass $M$ is defined by eq.~(\ref{M}). 
Imposing the condition
$\partial_\mu{\bf S}_{\mu\nu}=0$, one can further, similarly to Ref.~\cite{moresu3}, derive 
the desired two-point correlation function (string propagator):

\begin{equation}
\label{two}
\left<S_{\mu\nu}^\alpha(x)S_{\lambda\rho}^\beta(0)\right>=\delta^{\alpha\beta}
\varepsilon_{\mu\nu\gamma\sigma}\varepsilon_{\lambda\rho\xi\sigma}\frac{\zeta}{(M\Lambda^2)^2}
\partial_\gamma\partial_\xi(\partial^2-m^2)[D_0(x)-D_M(x)].
\end{equation}
It is known that, in this Gaussian approximation, all higher cumulants
vanish, i.e. the terms with $l\ge 2$ in eq.~(\ref{tens}) are
absent. Therefore, in this very dilute plasma approximation, the
Casimir-scaling law is preserved, since the terms that violate it
vanish.

\section{Summary}

In this paper, we have explored the $k$-string tension spectrum in the
SU($N$)-inspired 4-d dual Abelian--Higgs-type theory.  We have first
considered the London limit of this theory and demonstrated that, in
the leading semi-classical approximation when the small-sized closed
dual strings are completely disregarded, the $k$-string tension obeys
the Casimir-scaling law. In the same approximation, when closed
strings are disregarded, we have further explored the leading
correction to the Casimir scaling emerging due to the deviation from the
London limit, i.e.  due to the finiteness of the masses of the Higgs
bosons. This correction turns out to have the same $N$-ality
dependence as the correction, which one finds in the 3-d SU($N$)
Georgi--Glashow model when the non-diluteness of the monopole plasma
is taken into account. We have then addressed another type of
corrections to the Casimir scaling, which emerge in the London limit
when one accounts for the dilute plasma of closed dual strings. In the
leading low-density approximation, i.e. when the plasma is very
dilute, the respective correction is shown not to violate the Casimir
scaling. Instead, the correction of the next order in the
non-diluteness has the same $N$-ality dependence as the
above-mentioned correction emerging without closed strings in the
vicinity of the London limit. Finally, we have analyzed the
corrections that appear in higher orders in the non-diluteness of the
plasma of closed strings. Interestingly, the $1/N$ dependence of the
ratio $\sigma_k/\sigma_1$ does not satisfy the counting rules that were
spelled out for SU($N$) Yang--Mills theories in
Ref.~\cite{12}. However, one has to be careful in comparing our
results to the full non-Abelian gauge theory. In our model,
off-diagonal degrees of freedom are disregarded, the only remnant of
the non-Abelian structure being the quantization condition for the
magnetic charge. The discrepancy between the results presented here
(and in Ref.~\cite{jh}) and those obtained in Ref.~\cite{12} suggests
that a purely Abelian description of the SU($N$) vacuum is not
adequate to catch the full dynamics of the non-Abelian gauge theory,
in agreement with the conclusions
in~Refs.~\cite{Ambjorn:1998qp,Ambjorn:1999ym}. Further work along
these lines is needed to clarify this issue.

\acknowledgments Two of us (D.A. and D.E.) are grateful to Y.~Koma for
useful discussions. D.A. acknowledges the Alexander~von~Humboldt
Foundation for the financial support. He would also like to thank the
staff at the Institute of Physics of the Humboldt University of Berlin
for the kind hospitality.

\appendix
\section{A few technical points}

Let us first evaluate the effect produced by the determinant when
integrating over the ${\bf B}_\mu$-field in eq.~(\ref{gauss}). One has

\begin{equation}
\label{de}
\det{}^{-1/2}\left[\hat 1+\Xi\right]\simeq\exp\left[-\frac12{\rm Tr}{\,}\Xi+\frac14{\rm Tr}{\,}\Xi^2\right],
\end{equation}
where ``Tr'' includes both the trace over the indices $\alpha, \beta$
and the trace in the coordinate space.  Next terms have been omitted
since they describe self-interactions of the
$\varphi_i$-field, rather than the renormalization of $m$ and $m_H$,
which will be shown to be produced by the two retained terms.  The
first term on the r.h.s. of eq.~(\ref{de}) modifies eq.~(\ref{co}) as
follows:

$$\frac12(\partial_\mu\varphi_i)^2+\frac{m_H^2}{2}\varphi_i^2+\xi^{\alpha\beta}
\left(C^{\alpha\beta}+\frac{zm_H^4}{2}\delta^{\alpha\beta}\right)\longrightarrow
-\frac{4}{(\eta N)^2}\sum\limits_{i}^{}q_i^\alpha q_i^\beta q_i^\gamma q_i^\delta\times$$

$$\times\left[C^{\alpha\beta}(x)+\frac{zm_H^4}{2}\delta^{\alpha\beta}\right]
\int d^4x'
D_{m_H}(x-x')\left[C^{\gamma\delta}(x')+\frac{zm_H^4}{2}\delta^{\gamma\delta}\right],$$
where the regularization parameter $z$ is defined by the relation
$\left.\delta^{(4)}(x)\right|_{x=0}=zm_H^4$.  Note that, due to
eq.~(\ref{MH}) and the fact that $\left.\delta^{(4)}(x)\right|_{x=0}$
is $N$-independent, $z$ should scale with $N$ as ${\cal
O}(N^{-2})$. Apart from this requirement, $z$ can be chosen at
will. Up to an inessential constant addendum, the correction to the
action thus reads:

$$
-\frac{2zm_H^4}{(\eta N)^2}\sum\limits_{i}^{}q_i^\alpha q_i^\beta q_i^\gamma q_i^\delta
\int d^4xd^4x' D_{m_H}(x-x')\left[C^{\alpha\beta}(x)\delta^{\gamma\delta}+C^{\gamma\delta}(x')\delta^{\alpha\beta}
\right]=$$

$$=-\frac{2zm_H^2}{N\eta^2}\int d^4x C^{\alpha\alpha}.$$ Comparing
this result with the first term on the r.h.s. of eq.~(\ref{corr}), we
arrive at the following renormalization of $m^{-2}$:

\begin{equation}
\label{app1}
\frac{1}{m^2}\rightarrow\frac{1}{m^2}\left(1+\frac{2zm_H^2}{N\eta^2}\right),
\end{equation}
where the correction vanishes at large $N$ as $\frac{2zm_H^2}{N\eta^2}={\cal O}(N^{-1})$.
Let us now consider the second term on the r.h.s. of eq.~(\ref{de}). It modifies the mass term on the l.h.s. 
of eq.~(\ref{co}) as

$$
\frac{m_H^2}{2}\varphi_i^2\rightarrow\frac{m_H^2}{2}\sum\limits_{i,j}^{}\varphi_i\varphi_j\left[\delta_{ij}
-\frac{4zm_H^2}{(N\eta)^2}q_i^\alpha q_i^\beta q_j^\alpha q_j^\beta\right].$$
The tensor $q_i^\alpha q_i^\beta q_j^\alpha q_j^\beta$ should be proportional to $\delta_{ij}$, which is the only
tensor symmetric in indices $(i,j)$, and the proportionality coefficient is 1. The squared Higgs mass therefore
renormalizes as

\begin{equation}
\label{app2}
m_H^2\rightarrow m_H^2\left[1-\frac{4zm_H^2}{(N\eta)^2}\right].
\end{equation}
The obtained correction vanishes at large $N$:
$\frac{4zm_H^2}{(N\eta)^2}={\cal O}(N^{-2})$, so that $\frac{4zm_H^2}{(N\eta)^2}<1$ in the large-$N$ limit.
A more accurate condition on $z$, which should hold at any $N$, can be imposed by using 
the inequality of the London limit, eq.~(\ref{lam}). It yields:

$$
1>\frac{4zm_H^2}{(N\eta)^2}=\frac{16\lambda z}{N^2}\gg\frac{16 z}{N^2}\frac{(2\pi{\rm e}N)^2}{\bar\lambda}=
\frac{(8\pi{\rm e})^2z}{\bar\lambda},
$$ i.e. at a given $\bar\lambda$, $z$ should be chosen such that
$z\ll\frac{\bar\lambda}{(8\pi{\rm e})^2}$.  The squared
Landau--Ginzburg parameter then renormalizes according to
eqs.~(\ref{app1}), (\ref{app2}) as

$$
\kappa^2\rightarrow\kappa^2\left[1+\frac{2zm_H^2}{N\eta^2}\left(1-\frac{2}{N}\right)\right]=
\kappa^2\left\{1+{\cal O}(N^{-1})\cdot\left[1+{\cal
O}(N^{-1})\right]\right\}.$$ Therefore, upon renormalization, $\kappa$
remains large, and the respective correction vanishes in the large-$N$
limit.

Let us now evaluate the omitted $\Xi^2$-term on the r.h.s. of eq.~(\ref{gauss}). 
This term reads

\begin{equation}
\label{app4}
k_\mu^\alpha\xi^{\alpha\beta}\xi^{\beta\gamma}k_\mu^\gamma=\frac{8}{(\eta N)^2}k_\mu^\alpha k_\mu^\gamma
\varphi_i\varphi_j q_i^\alpha q_i^\beta q_j^\beta q_j^\gamma,
\end{equation}
with the monopole current ${\bf k}_\mu$ defined by
eq.~(\ref{ll1}). Since the tensors $k_\mu^\alpha k_\mu^\gamma$,
$\varphi_i\varphi_j$ are symmetric in indices $(\alpha,\gamma)$ and
$(i,j)$, respectively, it is natural to impose the following Ansatz:
$q_i^\alpha q_i^\beta q_j^\beta q_j^\gamma={\cal
N}\delta^{\alpha\gamma}\delta_{ij}$. The proportionality coefficient
${\cal N}$ can readily be found: ${\cal N}=\frac{1}{N-1}$, and the
term~(\ref{app4}) therefore reads $\frac{8{\bf k}_\mu^2}{(\eta
N)^2(N-1)}\varphi_i^2$. Furthermore, the characteristic amplitude of
the ${\bf k}_\mu$-field can be estimated by noticing that the
configuration of this field, which dominates in the partition
function, is the one at which each of the first two terms on the
r.h.s. of eq.~(\ref{higgs}) is of the order of unity. When applied to
the mass term, this requirement yields $L^2\sim |{\bf k}_\mu|^{-1}$,
where $L$ and $|{\bf k}_\mu|$ are the characteristic wavelength and
the amplitude of the ${\bf k}_\mu$-field, respectively. Applying
further the same requirement to the Coulomb interaction of monopole
currents, the second term on the r.h.s. of eq.~(\ref{higgs}), we have
$m^2|{\bf k}_\mu|^2L^6\sim 1$.  Substituting here the above
estimate for $L^2$, we obtain the desired estimate for the
characteristic value of the amplitude: $|{\bf k}_\mu|\sim m^2$. This
leads to the following estimate for the magnitude of the
term~(\ref{app4}): $\frac{m^4}{(\eta N)^2(N-1)}\varphi_i^2$. This term
therefore produces a small positive correction to $m_H$, whose
magnitude with respect to $m_H$ can be estimated as:

$$
\frac{m^2}{\eta N\sqrt{N-1}}\frac{1}{m_H}={\cal O}\left(\frac{1}{\kappa\sqrt{N}}\right).$$

\section{More on Casimir scaling in the 3-d SU($N$) Georgi-Glashow model}

It has been proved in~\cite{jh} that, in the 3-d SU($N$) Georgi-Glashow model, 
Casimir scaling holds for an {\it arbitrarily shaped} surface (i.e. the world sheet of a $k$-string), 
provided that 
the density of monopole plasma is much lower than the mean one. The purpose of this Appendix is to
show that, in case of a flat surface, Casimir scaling in this model is an {\it exact} result, 
not requiring the condition that the density of the monopole plasma is much lower than the mean one.
To this end, let us consider the confining part of the $k$-th power of the fundamental Wilson loop,
i.e. the part produced by monopoles. It has the form $\left<W_k(C)\right>_{\rm mon}=\sum\limits_{{a_1,\ldots,
a_k=1}\choose{{\rm with}~ {\rm possible}~ {\rm
coincidences}}}^{N}W_{a_1,\ldots, a_k}(C)$, where 

$$
W_{a_1,\ldots, a_k}(C)=\frac{1}{{\mathcal Z}_{\rm mon}}
\int{\mathcal D}{\bf B}_\mu\delta\left(\varepsilon_{\mu\nu\lambda}\partial_\nu{\bf B}_\lambda\right)
\int {\mathcal D}{\bf l}
\exp\Biggl\{\int d^3x\Biggl[-\frac{g_m^2}{2}{\bf B}_\mu^2+$$

\begin{equation}
\label{Waa}
+ig_m{\bf l}\partial_\mu{\bf B}_\mu+
2\zeta\sum\limits_{i}^{}
\cos\left(g_m{\bf q}_i{\bf l}\right)\Biggr]+
4\pi i{\bf M}_k^{(n)}\int\limits_{\Sigma(C)}^{}d\sigma_\mu{\bf B}_\mu\Biggr\},
\end{equation}
where ${\mathcal Z}_{\rm mon}$ is the same functional integral, but with the last term [which describes the 
flux of the magnetic field through an arbitrary surface $\Sigma(C)$] set equal to zero.
The constraint $\varepsilon_{\mu\nu\lambda}\partial_\nu{\bf B}_\lambda=0$ 
imposes the fact that free photons, inessential for confinement, are not taken into account.
The dimensionalities of the magnetic coupling constant, $g_m$, dual photon field, ${\bf l}$, and the 
(exponentially small) monopole fugacity $\zeta$ are $[{\rm mass}]^{-1/2}$, $[{\rm mass}]^{1/2}$, and 
$[{\rm mass}]^3$, respectively; for more details on the model and eq.~(\ref{Waa}) at $k=1$ see~\cite{jh, cris}.
In eq.~(\ref{Waa}), ${\bf M}_k^{(n)}$ again denotes the sum $\sum\limits_{i=1}^{k}{\bf m}_{a_i}$, where some $n$
indices out of $k$ can now coincide. 

For the contour $C$ located in the $(x,y)$-plane,
the saddle-point equations stemming from eq.~(\ref{Waa}) read

\begin{equation}
\label{spA1}
ig_m {\bf l}'+g_m^2{\bf B}-4\pi i{\bf M}_k^{(n)}\delta(z)=0,
\end{equation}

\begin{equation}
\label{spA2}
i{\bf B}'-2\zeta\sum\limits_{i}^{}{\bf q}_i\sin(g_m{\bf q}_i{\bf l})=0,
\end{equation}
where $'\equiv d/dz$, and the natural Ansatz ${\bf B}_\mu=\delta_{\mu 3}{\bf B}(z)$, ${\bf l}={\bf l}(z)$
has been adopted. 
Next, as it follows from eq.~(\ref{spA1}), ${\bf B}\propto i{\bf M}_k^{(n)}$, therefore 
$W_{a_1,\ldots, a_k}(C)\to{\rm e}^{-\left({\bf M}_k^{(n)}\right)^2\sigma|\Sigma(C)|}$ at $|\Sigma(C)|\to\infty$, 
where the string tension $\sigma$
is $k$-independent. Since 
$\left({\bf M}_k^{(n)}\right)^2=C_k+\frac{n^2-n}{2}$, in the 
limit of asymptotically large areas $|\Sigma(C)|$ of interest, we arrive at a Feynman-Kac--type formula,

\begin{equation}
\label{FK}
\left<W_k(C)\right>_{\rm mon}={\rm e}^{-C_k\sigma|\Sigma(C)|}\sum\limits_{n=1}^{k}c_n
{\rm e}^{-\frac{n^2-n}{2}\sigma|\Sigma(C)|},
\end{equation}
with some positive coefficients $c_n$; therefore, only the case $n=0$ is relevant in eq.~(\ref{Waa}) 
(cf. ref.~\cite{jh}).
We should, thus, solve the system of eqs.~(\ref{spA1}), (\ref{spA2}) with ${\bf M}_k^{(0)}$, which coincide 
with ${\bf M}_k$ from the main text. Setting ${\bf B}(z)={\bf M}_kB(z)$, 
${\bf l}(z)={\bf M}_kl(z)$, we see that eq.~(\ref{spA1}) takes the same form as in the fundamental case, namely

\begin{equation}
\label{neweq1}
ig_ml'+g_m^2B=4\pi i\delta(z),
\end{equation} 
whereas to handle eq.~(\ref{spA2}), one should notice that 
any root vector is a difference of two weight vectors, ${\bf q}_i\equiv {\bf q}_{ab}={\bf m}_a-{\bf m}_b$.
In particular, positive roots, we are dealing with, are those with $b<a$, therefore eq.~(\ref{spA2}) takes the form

$$
\sum\limits_{b<a}^{}{\bf M}_k({\bf m}_a-{\bf m}_b)\sin\left[g_m{\bf M}_k({\bf m}_a-{\bf m}_b)l\right]=
\frac{i}{2\zeta}C_kB'.
$$
Using the symmetry of the expression under the sum on the l.h.s. with respect to $a\leftrightarrow b$, we can 
rewrite this equation as 

\begin{equation}
\label{SPEQ}
\sum\limits_{a,b=1}^{N}{\bf M}_k{\bf m}_a\left[\sin(g_m{\bf M}_k{\bf m}_al)\cos(g_m{\bf M}_k{\bf m}_bl)-
\cos(g_m{\bf M}_k{\bf m}_al)\sin(g_m{\bf M}_k{\bf m}_bl)\right]=\frac{i}{2\zeta}C_kB'. 
\end{equation}
We should further perform the four sums on the l.h.s.; let us begin with the first one, $\sum\limits_{a=1}^{N}
({\bf M}_k{\bf m}_a)\sin(g_m{\bf M}_k{\bf m}_al)$. Appartently, there are $k$ terms in this sum, for which 
${\bf m}_a$ coincides with some of the $k$ weight vectors, which enter ${\bf M}_k$. Using the relation
${\bf m}_a{\bf m}_b=(\delta_{ab}-N^{-1})/2$, we have for such terms ${\bf M}_k{\bf m}_a=\frac{N-1}{2N}-(k-1)\frac{1}{2N}=
\frac{N-k}{2N}$. For the other $(N-k)$ terms in the sum, ${\bf m}_a$ does not coincide with any weight vector 
in ${\bf M}_k$, hence ${\bf M}_k{\bf m}_a=-\frac{k}{2N}$ for such terms. We, therefore, obtain

$$\sum\limits_{a=1}^{N}
({\bf M}_k{\bf m}_a)\sin(g_m{\bf M}_k{\bf m}_al)=k\cdot\frac{N-k}{2N}\sin\left(g_m\frac{N-k}{2N}l\right)+
(N-k)\cdot\frac{k}{2N}\sin\left(g_m\frac{k}{2N}l\right)=$$

$$=C_k\left[\sin\left(g_m\frac{k}{2N}l\right)+
\sin\left(g_m\frac{N-k}{2N}l\right)\right].$$
Remarkably, already this expression alone is manifestly invariant under $k\leftrightarrow (N-k)$. In the same way,
we obtain for the three other sums the following expressions:

$$
\sum\limits_{b=1}^{N}\cos(g_m{\bf M}_k{\bf m}_bl)=k\cos\left(g_m\frac{N-k}{2N}l\right)+
(N-k)\cos\left(g_m\frac{k}{2N}l\right),
$$

$$\sum\limits_{a=1}^{N}
({\bf M}_k{\bf m}_a)\cos(g_m{\bf M}_k{\bf m}_al)=C_k\left[\cos\left(g_m\frac{N-k}{2N}l\right)-
\cos\left(g_m\frac{k}{2N}l\right)\right],$$

$$
\sum\limits_{b=1}^{N}\sin(g_m{\bf M}_k{\bf m}_bl)=k\sin\left(g_m\frac{N-k}{2N}l\right)-
(N-k)\sin\left(g_m\frac{k}{2N}l\right).
$$
The l.h.s. of eq.~(\ref{SPEQ}) then takes the form  $C_kN\sin\frac{g_ml}{2}$, and the whole equation becomes

\begin{equation}
\label{neweq2}
B'+2i\zeta N\sin\frac{g_ml}{2}=0,
\end{equation}
that also coincides with that of the fundamental case~\cite{cris}. The solution to the system of equations~(\ref{neweq1}), 
(\ref{neweq2}) reads

\begin{equation}
\label{solution}
B(z)=i\frac{8m_D}{g_m^2}\frac{{\rm e}^{-m_D|z|}}{1+{\rm e}^{-2m_D|z|}},~~
l(z)=\frac{8}{g_m}{\,}{\rm sgn}{\,}z\cdot\arctan\left({\rm e}^{-m_D|z|}\right),
\end{equation}
where $m_D=g_m\sqrt{N\zeta}$ is the Debye mass of the dual photon ${\bf l}$. [This mass is visible in 
eq.~(\ref{Waa}) with $C=0$, where the ${\bf B}_\mu$-field is integrated out to produce the kinetic term of the 
${\bf l}$-field, and cosine is expanded up to the quadratic term.] Inserting the so-obtained field $B(z)$
into eq.~(\ref{Waa}), we obtain $\sigma=c g\sqrt{N\zeta}$, where $g=4\pi/g_m$. The proportionality coefficient $c$ here depends 
on the range of average of $B(z)$, since, according to eqs.~(\ref{solution}), the string is exponentially thick, 
$|z|\lesssim m_D^{-1}$. For instance, if we simply choose the value $B(0)$, then $c=4$. [For comparison, the value of $c$
one obtains in the case when the density of the monopole plasma is much lower
than the mean one, $\zeta N(N-1)$, is~\cite{cris}
$\pi$, that can be shown to approximately correspond to the following range of average: 
$|z|<\frac{\sqrt{6-\frac{3\pi}{2}}}{m_D}$.] The obtained string tension is manifestly $k$-independent.

Equation~(\ref{FK}) then leads to the conclusion that, for a flat surface, Casimir scaling holds in the full sine-Gordon
theory describing the monopole plasma in the 3-d Georgi-Glashow model. In another words, for a flat surface, Casimir 
scaling holds also at monopole densities close to the mean one, $\zeta N(N-1)$, rather than only at the densities much smaller
than $\zeta N(N-1)$, as it takes place for a non-flat 
surface~\cite{jh}~\footnote{It has been argued in ref.~\cite{jh} that non-diluteness corrections are suppressed if $N\lesssim {\cal O}
\left({\rm e}^{S_0/2}\right)$. Here, $S_0=\frac{4\pi\epsilon m_W}{g^2}$ is a single-monopole action, $S_0\gg 1$,
where $m_W$ is the W-boson mass, and the function $\epsilon$ describes quantum corrections to the classical 
expression, $1\le\epsilon <1.8$. Although this boundary on $N$ (above which non-diluteness effects might significantly distort
Casimir scaling) is exponentially large, it nevertheless does exist for a non-flat surface. Instead, as we have just seen,   
for a flat surface, Casimir scaling holds at any, whatever large, $N$.}. It is finally reasonable to have some feeling 
on how much a surface should deviate from a flat one in order that the Casimir scaling starts violating, if the density
of monopoles is the mean one, $\zeta N(N-1)$.
To this end, let us consider a straight string (apparently corresponding to a flat surface) 
of a minimal possible length, $m_D^{-1}$. According to the first of eqs.~(\ref{solution}),
the field $B(z)$ at the end points of such a string decreases above and below the surface also at the distance $m_D^{-1}$.
Therefore, if we start bending the string such that it forms a piece of a circle, 
the two solutions overlap with each other if the radius 
of this circle is $\le m_D^{-1}$. The critical situation when this radius is equal to $m_D^{-1}$ corresponds to
the distance between the end points of the string equal to $2m_D^{-1}\sin\frac12$. This is the minimal
distance which should hold between two points, separated by the distance $m_D^{-1}$ along the string, at which 
the surface can still be considered as flat from the point of view of the Casimir scaling.

\end{document}